# Observation of the Josephson effect in Pb/Ba$_{1-x}$K$_x$Fe$_2$As$_2$ single crystal junctions

*Xiaohang Zhang[1], Yoon Seok Oh[2], Yong Liu[2], Liqin Yan[2], Kee Hoon Kim[2], Richard L. Greene[1] & Ichiro Takeuchi[3]*

[1]CNAM and Department of Physics, University of Maryland, College Park, Maryland 20742, USA

[2]CSCMR & FPRD, Department of Physics and Astronomy, Seoul National University, Seoul 151-747, Republic of Korea

[3]Department of Materials Science and Engineering, University of Maryland, College Park, Maryland 20742, USA

**ABSTRACT** We have fabricated *c*-axis Josephson junctions on single crystals of Ba$_{1-x}$K$_x$Fe$_2$As$_2$ by using Pb as the counter electrode in two geometries, planar and point contact. Junctions in both geometries show resistively shunted junction *I-V* curves below the $T_C$ of the counter electrode. Microwave induced steps were observed in the *I-V* curves, and the critical currents are suppressed with an in-plane magnetic field with well-defined modulation periods indicating that the Josephson current is flowing in a manner consistent with the small to intermediate sized junction limit. $I_C R_N$ products of up to 0.3 mV have been observed in these junctions at 4.2 K. The observation of Josephson coupling along the *c*-axis between Ba$_{1-x}$K$_x$Fe$_2$As$_2$ and a conventional superconductor suggests the existence of a *s*-wave symmetry in this class of iron pnictide superconductors.

The iron pnictides are a new class of transition-metal based superconductors which have produced great interest since their discovery about one year ago [1-7]. A central issue for understanding the origin of the superconductivity in these materials is the pairing symmetry. Theoretically, a variety of pairing symmetries have been proposed [8-10]. Experimentally, the pairing has been interpreted to be nodeless [11-17] or nodal [17-20]. Phase sensitive experiments, such as the Josephson effect, the grain boundary effect, and the SQUID interferometry, have played an important role in establishing the pairing symmetry of cuprates [21,22]. In particular, a half-integer flux-quantum is related to internal $\pi$ phase shifts in the order parameter and has been observed in cuprates with *d*-wave symmetries [23]. A scanning SQUID microscopy measurement on polycrystalline samples of the $NdFeAsO_{1-x}F_x$ (1111) pnictides indicated the likely absence of a half-integer flux-quantum effect [24] associated with Josephson tunneling across grain boundaries. The absence of the effect in the 1111 iron pnictides suggests that *d*-wave symmetry may not be the pairing nature in this class of pnictides. To date, no phase-sensitive experiments have been performed on single crystals of iron pnictide superconductors with the $BaFe_2As_2$ (122) structure.

For a Josephson junction, the value of the maximum supercurrent is given by $I_S = I_C \sin\gamma$, where $I_C$ is the junction critical current, and $\gamma$ is the gauge-invariant phase difference across the junction. For junctions between two single phase *s*-wave superconductors, $\gamma$ is the relative phase between the two superconductors at zero field. More generally, the magnitude and the phase of the order parameters need also be taken into account for the Josephson effect [22], and cancellation of the Josephson current is expected in certain junctions that consist of one or two non-*s*-wave superconductors. In particular, Josephson coupling between a superconductor with an unknown symmetry and a *s*-wave superconductor can be used to test the non-vanishing term in the order parameter along a certain direction.

In this letter, we report the first strong evidence for Josephson coupling between a conventional BCS superconductor and a single crystal of $Ba_{1-x}K_xFe_2As_2$ along the *c*-axis. Microwave induced steps were observed in the *I-V* curves of the Josephson junctions, and the critical currents are suppressed with an in-plane magnetic field in a Fraunhofer-like manner consistent with the small to intermediate junction limit. The observation of the Josephson coupling along the *c*-axis of the iron pnictide superconductor suggests the existence of an *s*-wave superconducting order parameter in the 122 class of iron pnictide superconductors. Furthermore, within the *s*±-wave pairing symmetry [9], our results suggest the predominance of one sign in the magnitude of the order parameter.

Two batches of *c*-plane $Ba_{1-x}K_xFe_2As_2$ single crystals used in this study were grown using the Sn flux-method [25,26]. Following the synthesis, the compositions were determined by Energy-dispersive X-ray spectroscopy (EDX), and the potassium concentrations were found to be $x = 0.49$ and $0.29$. The corresponding bulk transition temperatures of the single crystals with the two doping concentrations were determined to be about 26 K and 29 K, respectively, consistent with the fact that the crystals were not optimally doped. The transition widths were less than 2 K. Prior to the junction fabrication, no special cleaning/treatment procedure was performed on the surface of the crystals. Scanning electron microscopy (SEM) indicates that the surfaces of the crystals consist of large flat terrace-free regions up to $(mm)^2$ in size. As an example, an SEM image of a single crystal is shown in Fig. 1a. Both planar geometry and point contact geometry were used in junction fabrication. In the planar junction geometry, epoxy was used to define a gap on the surface of the crystal. A 30 nm Ag layer was first thermally evaporated across the gap through a shadow mask, followed by deposition of a PbIn (with ≈ 5 wt. % In) layer with a thickness of 500 nm. The planar junctions had a defined size of about $200 \times 300$ $(\mu m)^2$. In the point contact geometry, a sharpened Pb tip was directly pressed against the surface of the single crystal at the liquid helium temperature. In the latter geometry, the flattened tip area was found to be typically in the range of 100 × 100 (μm)$^2$ to 200 × 200 (μm)$^2$ following the pressing contact. Several point contact junctions and a planar junction were measured, and they all showed similar behavior.

Fig. 1b shows the temperature dependence of the zero bias junction resistance $R_J$ measured form the planar junction with an ac current modulation of 100 μA. At temperatures slightly higher than the transition temperature of PbIn, Andreev reflection (AR) behavior was observed in conductance versus voltage (not shown here), indicating the presence of a highly transparent interface in the junction. Two observations on the $R_J$(T) curve indicate that our junctions are indeed in the AR regime when the temperature is between the two superconducting transition temperatures (~ 7 K for PbIn, and ~ 20 K for the pnictide in this sample): first, consistent with the AR theory of a superconductor/normal metal (SN) junction, the zero bias junction resistance gets suppressed approximately 50% in going from 20 K to 7 K; secondly, the decrease of the junction resistance occurs in a broad temperature range rather than a narrow range slightly below the $T_C$ of the crystal, suggesting the reduction in the junction resistance is predominantly from the interface, rather than from the bulk single crystal, which by itself shows a sharper resistance drop upon undergoing superconducting transition.

Below the superconducting temperature of the Pb (PbIn) counter electrode, a robust supercurrent was consistently observed in both types of junctions. Fig. 2a shows a typical *I-V* characteristic of the junctions. The *I-V* curve shows no significant hysteresis and can be well described by the resistively shunted junction (RSJ) model [27] (Fig. 2b). Sharp Shapiro steps are clearly observed at voltages corresponding to multiples of *hf*/2*e* with irradiation of a 4 GHz microwave field, confirming the flow of Cooper pairs through the interface (Fig. 2c). Upon increasing the power of the microwave field, the critical current was found to be completely suppressible. The $I_CR_N$ products (where $R_N$ is the normal state junction resistance) were found to range from 3 µV to 300 µV at 4.2 K. The temperature dependence of the $I_CR_N$ product of the planar junction is shown in Fig. 2d. The linear temperature dependence of the $I_CR_N$ down to $0.3T_C$ is consistent with the de Gennes' dirty limit theory [28] as indicated by the solid curve in the figure.

Upon application of the magnetic field in the in-plane direction, junctions showed relatively symmetric Fraunhofer-like modulation patterns of the critical currents with well-defined repeating periods indicating that our devices are in the small to intermediate junction limit with fairly uniform current flow through confined active areas located somewhere within the defined junction area for the planar geometry or within the flattened tip area for the point contact geometry. As an example, the magnetic modulation pattern of the critical current of a point contact junction is shown in Fig. 3. The solid curve is a model fit which takes into account the details of the current flow within this effective junction area [29,30]. Using the penetration depths of 50 nm for Pb and 200 nm for $Ba_{1-x}K_xFe_2As_2$, the observed main modulation period corresponds to the effective junction width $W$ of ≈ 14 µm. We note that compared to the flattened tip contact area determined after the experiment (~ 100 × 100 (µm)$^2$), this suggests the actual active junction region is much smaller. From this actual size of the active area, the Josephson critical current density, $J_C$, is estimated to be about $1.0×10^7$ A/m$^2$, which in turn places the Josephson penetration depth, $\lambda_J$, to be 10 µm. Therefore, we find that $W/\lambda_J = 1.4$ for this junction. The value of $W/\lambda_J$ reflects the relative strength of the self-field effect compared to the external field and provides an indication as to where the junction lies between the small junction and the large junction limits [29,30]. In this case, the obtained value of 1.4 is consistent with the observed shape of the magnetic diffraction pattern being in the intermediate-sized regime. Further local fluctuation within the junction active area might be the reason for the remaining small fraction (≈ 5%) of the current which is not modulated in this junction.

A number of junctions were studied, and the actual active junction area where Josephson current is flowing was always found to be of the order of ~ 10 × 10 (μm)$^2$. This indicates that over a much larger area (≥ 100 × 100 (μm)$^2$), there exists a distribution of superconducting and non-superconducting regions. When the junctions were fabricated (and contacts were made) on the crystal surfaces, there were only these 10 × 10 (μm)$^2$ order sized pockets where the very surface was still superconducting. The density of such pockets appears to be such that we are able to "hit" one or two over a ~ 100 × 100 (μm)$^2$ area. The changing chemistry and subsequent loss of superconductivity at the very surface of 122 crystals following synthesis due to oxidation of As or K terminated surfaces in ambient atmosphere are known. The extent of this oxidation is such that after a period of a month or so, a crystal kept in ambient seems to lose superconductivity at the surface [31].

The typical junction resistance observed here is of the order of 10 mΩ. If we were to "average" this over the nominal junction size of ~ 100 × 100 (μm)$^2$, the apparent interface resistivity would be of the order of $10^{-6}$ Ωcm$^2$ or larger. Previous SN interface studies with cuprates and a normal metal have shown that such a large interface resistivity cannot carry Josephson coupling due to the lack of interface transparency [32]. Instead, if we use ~ 10 × 10 (μm)$^2$ as the typical pocket area (and assuming the rest of the area to be insulating), the interface resistivity value is reduced to the range (≤ $10^{-8}$ Ωcm$^2$) where Josephson coupling becomes much more favorable. This indicates that the observed junction resistance is also mostly from the small active junction area, and confirms our simple model that the transport is almost exclusively taking place only at the effective interface region where the surface of the crystal is superconducting. The exact nature and the thickness of the "dead layer" at the surface in the surrounding areas formed due to oxidation are not known at this time, but this indicates that such a layer is generally insulating and too thick for any transport or Josephson coupling to take place.

Similar to the case of an cuprate/normal metal SN interface fabricated with a relatively pristine cuprate surface [32], the present pnictide/Ag interface (at the $\sim 10 \times 10$ $(\mu m)^2$ superconducting region) in the planar junction may also be modeled as consisting of nanometer seized arrays of filamentary conduction paths or having a very thin dielectric tunneling layer, which accounts for the residual interface resistance. Structural details of point contact junctions are often unpredictable, but again in the present junctions, we believe the transport properties are dominated by the microstructural state of the active junction region on the pnictide side. Thus, despite the difference in the geometries, both types of junctions can behave in a similar manner.

Comparing the $\sim 10 \times 10$ $(\mu m)^2$ sized active junction areas with the above mentioned large terrace-free flat surface areas on which the junctions were made, we conclude that our Josephson current is flowing directly along the *c*-axis direction, and *not* through the *ab*-plane via the terrace ledges.

The observation of the Josephson coupling between a conventional *s*-wave superconductor and an iron pnictide single crystal along the *c*-axis indicates the pairing symmetry is not a spin-triplet *p*-wave as the case in $Sr_2RuO_4$ [33]. Although the present observation does not necessarily exclude a *d*-wave pairing symmetry, it suggests the presence of a non-vanishing order parameter along the *c*-axis. In fact, similar observation of supercurrent along the *c*-axis of $YBa_2Cu_3O_{7-x}$ in tunneling Josephson junctions [34] had been a puzzle because of its inconsistency with the *d*-wave symmetry clearly revealed in other phase-sensitive experiments [35]; The dilemma was then resolved through *c*-axis twin boundary Josephson tunneling experiments [36], which indicated a mixed pairing state in $YBa_2Cu_3O_{7-x}$ with a dominant *d*-wave and a subdominant *s*-wave component.

In this work, the non-vanishing supercurrent term along the *c*-axis of $Ba_{1-x}K_xFe_2As_2$ suggests a possible *s*-wave pairing symmetry of the non-vanishing order parameter. As discussed above, a prevalent proposal of the pairing symmetry for iron pnictides is *s*±-wave with sign changes between Fermi pockets [9,10]. In this scenario, the phase of the symmetry is pinned with Fermi pockets, thus the carrier type. Due to doping, one carrier type can possess larger Fermi pockets; therefore, the Fermi velocity will be larger, and the magnitude of the order parameter with the corresponded sign will be dominant and both would lead to a non-vanishing order parameter term along the *c*-axis.

In summary, we observed robust Josephson coupling in Pb/ $Ba_{1-x}K_xFe_2As_2$ junctions along the *c*-axis of single crystals in both planar and point-contact geometries. RSJ type *I-V* characteristics were found. The magnetic diffraction pattern indicates the observed Josephson current is flowing mainly through active small junction areas which are highly transparent. The observation of the *c*-axis Josephson effect in this 122 iron pnictides rules out a *pure p*-wave or *d*-wave pairing symmetry in these materials; on the other hand, the result can be explained within the proposed *s*±-wave pairing symmetry. Moreover, the success in performing Josephson effect measurements on iron pnictides opens the door for further symmetry-sensitive or phase-sensitive investigations as well as exploration of device applications.

The authors would like to thank I. I. Mazin, J. Paglione, and F. Wellstood for fruitful discussions, and J. S. Kim for acquiring EDX data. X. Z. acknowledges N. Butch, K. Jin, S. Saha, P. Bach, A. Luykx, and D. Kan for technical help and useful discussions. X. Z. and R. L. G. are supported by the NSF under DMR-0653535 and I. T. is supported by NSF MRSEC at UMD (DMR 0520471). The work at SNU was supported by the National Research Lab program (M10600000238).

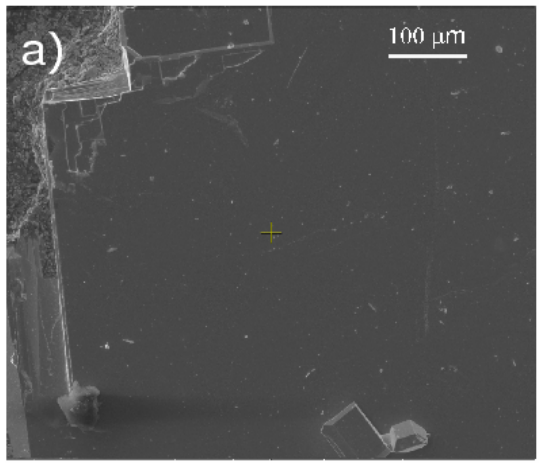 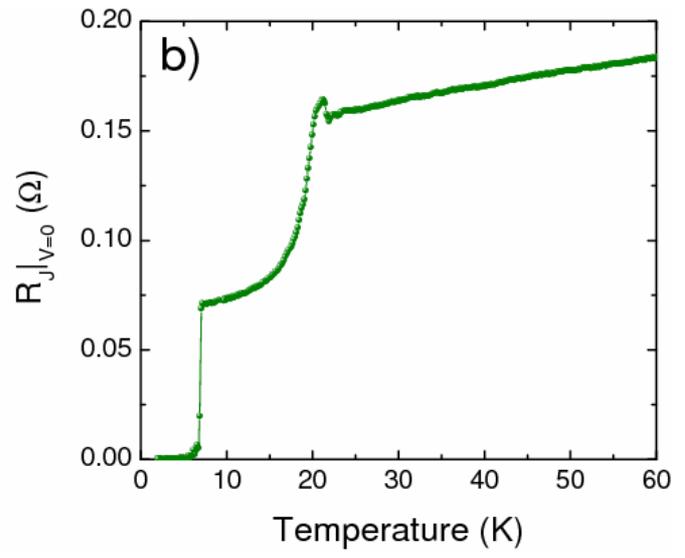

FIG. 1 (color online) a) An SEM image of a single crystal with a large flat surface area; b) Temperature dependence of the junction resistance for the planar junction.

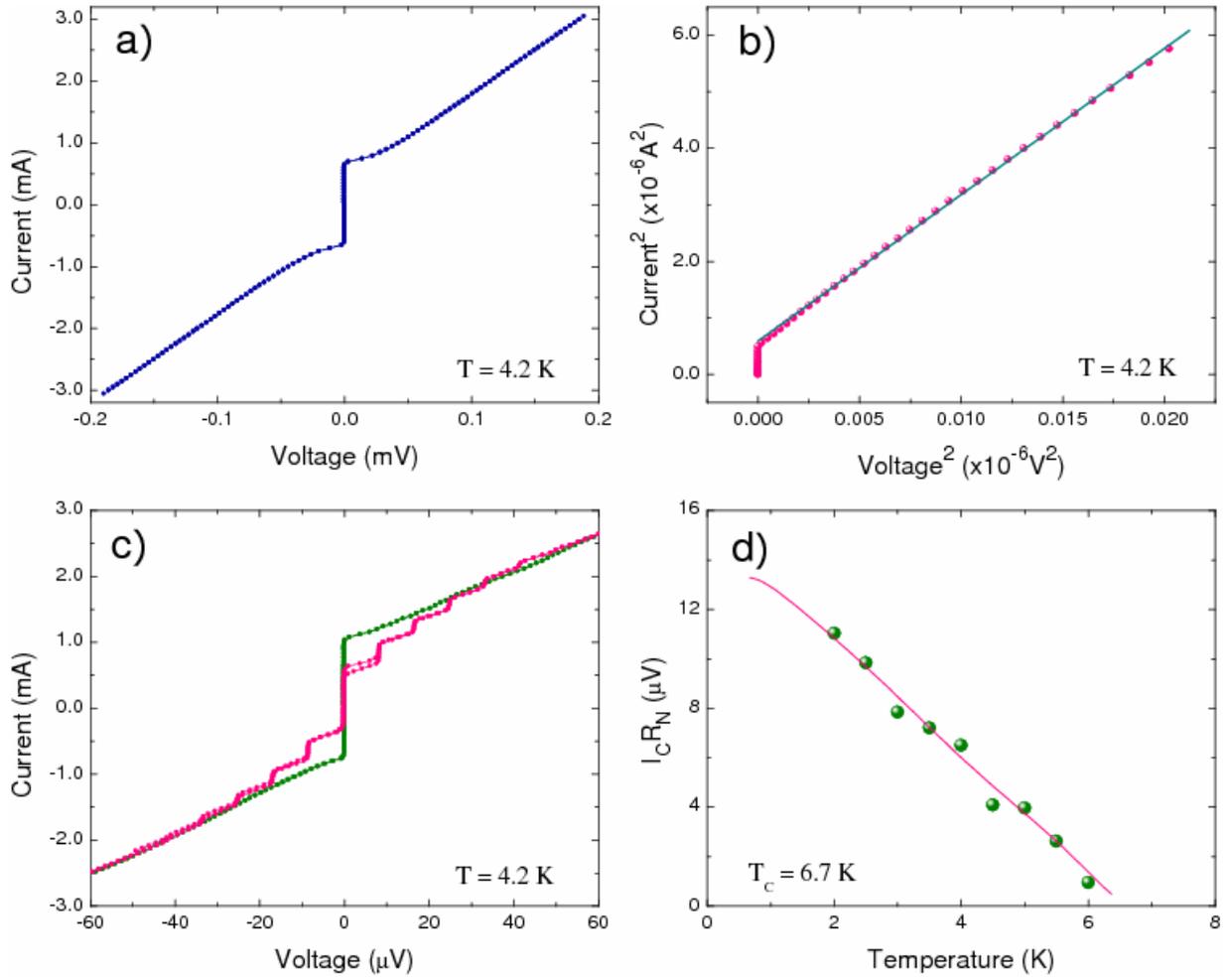

FIG. 2 (color online) a) *I-V* characteristic of a point contact junction; b) fit of the *I-V* curve in a) to the RSJ model: dots are experimental data and solid line is the fit; c) *I-V* characteristics of a point contact junction with (pink dots) or without (green dots) application of 4 GHz microwave irradiation; d) $I_C R_N$ products of the planar junction as function of temperature. The solid curve is a fit based on de Gennes's dirty limit theory.

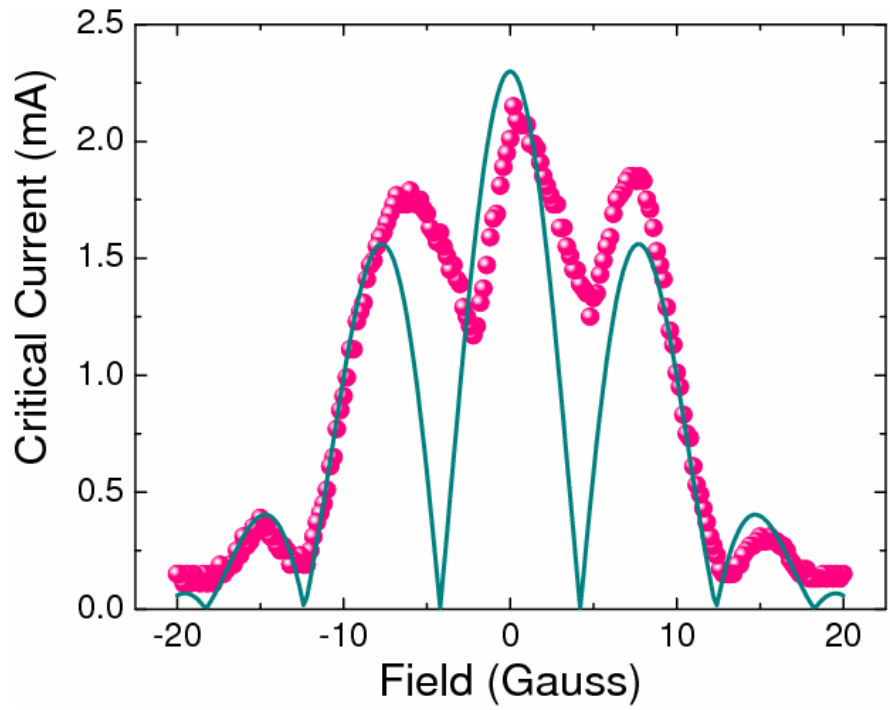

FIG. 3 (color online) Magnetic diffraction pattern of a point contact junction. Solid line is a model fit to the data based on the model described in the text.

REFERENCES

[1] Y. Kamihara, T. Watanabe, M. Hirano, and H. Hosono, J. Am. Chem. Soc. **130**, 3296 (2008).

[2] G. F. Chen *et al.*, Phys. Rev. Lett. **101**, 057007 (2008).

[3] Z. A. Ren *et al.*, Chin. Phys. Lett. **25**, 2215 (2008).

[4] X. H. Chen *et al.*, Nature **453**, 761 (2008).

[5] H. H. Wen *et al.*, Europhys. Lett. **82**, 17009 (2008).

[6] A. S. Sefat *et al*., Phys. Rev. B **77**, 174503 (2008).

[7] M. Rotter, M. Tegel, and D. Johrendt, Phys. Rev. Lett. **101**, 107006 (2008).

[8] X. Dai, Z. Fang, Y. Zhou, and F. C. Zhang, Phys. Rev. Lett. **101**, 057008 (2008); P. A. Lee and X. G. Wen, Phys. Rev. B **78**, 144517 (2008); Q. Si and E. Abrahams, Phys. Rev. Lett. **101**, 076401 (2008).

[9] I. I. Mazin, D. J. Singh, M. D. Johannes, and M. H. Du, Phys. Rev. Lett. **101**, 057003 (2008).

[10] V. Cvetkovic and Z. Tesanovic, arXiv:0804.4678; K. Seo, B. A. Bernevig, and J. Hu, Phys. Rev. Lett. **101**, 206404 (2008); A. V. Chubukov, D. V. Efremov, and I. Eremin, Phys. Rev. B, 78, 134512 (2008).

[11] T. Y. Chen *et al*., Nature **453**, 1224 (2008).

[12] H. Ding *et al.*, Europhys. Lett. **83**, 7001 (2008).

[13] T. Kondo *et al.*, Phys. Rev. Lett. **101**, 147003 (2008).

[14] G. Li *et al.*, Phys. Rev. Lett. **101**, 107004 (2008).

[15] L. Malone *et al.*, arXiv:0806.3908.

[16] C. Martin *et al.*, arXiv:0807.0876.

[17] K. A. Yates *et al.*, arXiv:0812.0977.

[18] L. Shan *et al.*, Europhys. Lett. **83**, 57004 (2008).

[19] O. Millo *et al.*, Phys. Rev. B **78**, 092505 (2008).

[20] M. C. Boyer *et al.*, arXiv:0806.4400.

[21] D. J. van Harlingen, Rev. Mod. Phys. **67**, 515 (1995).

[22] C. C. Tsuei and J. R. Kirtley, Rev. Mod. Phys. **72**, 969 (2000).

[23] C. C. Tsuei *et al.*, Phys. Rev. Lett. **73**, 593 (1994).

[24] Hicks C. W. *et al.*, arXiv:0807.0467v2.

[25] N. Ni *et al.*, Phys. Rev. B **78**, 014507 (2008).

[26] H. J. Kim *et al.*, arXiv:0810.3186v1.

[27] W. C. Stewart, Appl. Phys. Lett. **12**, 277 (1968); D. E. McCumber, J. Appl. Phys. **39**, 3113 (1968).

[28] P. G. de Gennes, Rev. Mod. Phys. **36**, 225 (1964).

[29] I. Takeuchi *et al.*, Appl. Phys. Lett. **68**, 1564 (1996).

[30] A. Barone and G. Paterno, *Physics and Applications of the Josephson Effect* (Wiley, New York, 1982).

[31] Preliminary low temperature STM experiments performed on the crystals over a period of a month or so have indicated degradation in conductivity at the surface over the period during which the crystals were kept in ambient atmosphere.

[32] I. Takeuchi *et al.*, IEEE Trans. Magn. **27**, 1626 (1991)

[33] A. P. Mackenzie and Y. Maeno, Rev. Mod. Phys. **75**, 657 (2003).

[34] A. G. Sun *et al.*, Phys. Rev. Lett. **72**, 2267 (1994); A. G. Sun *et al.*, Phys. Rev. B **54**, 6734 (1996).

[35] D. A. Wollman *et al.*, Phys. Rev. Lett. **71**, 2134 (1993); D. A. Wollman *et al.*, Phys. Rev. Lett. **74**, 797 (1995).

[36] K. A. Kouznetsov *et al.*, Phys. Rev. Lett. **79**, 3050 (1997).